\documentclass[aps,pra,twocolumn,amsmath,amssymb,nofootinbib,superscriptaddress]{revtex4}

\newcommand{\bra}[1]{\langle#1|}
\newcommand{\ket}[1]{|#1\rangle}

\newcommand{\expec}[1]{\langle #1\rangle}

\usepackage[pdftex]{graphicx}
\usepackage{mathrsfs}
\usepackage[usenames,dvipsnames]{color}
\usepackage[colorlinks = true, 
linkcolor = red, urlcolor = black, citecolor = red]{hyperref}
\usepackage{comment}
\usepackage{amsmath, amssymb}
\usepackage{mathtools}
\usepackage{notes2bib}
\usepackage{float}
\begin{document}

\bibliographystyle{apsrev}

\title{Linear Optical Quantum Metrology with Single Photons ---\\ Exploiting Spontaneously Generated Entanglement to Beat the Shot-Noise Limit}

\author{Keith R. Motes}
\email[]{motesk@gmail.com}
\affiliation{Centre for Engineered Quantum Systems, Department of Physics and Astronomy, Macquarie University, Sydney NSW 2113, Australia}

\author{Jonathan P. Olson}
\affiliation{Hearne Institute for Theoretical Physics and Department of Physics \& Astronomy, Louisiana State University, Baton Rouge, LA 70803}

\author{Evan J. Rabeaux}
\affiliation{Hearne Institute for Theoretical Physics and Department of Physics \& Astronomy, Louisiana State University, Baton Rouge, LA 70803}

\author{Jonathan P. Dowling}
\affiliation{Hearne Institute for Theoretical Physics and Department of Physics \& Astronomy, Louisiana State University, Baton Rouge, LA 70803}

\author{S. Jay Olson}
\affiliation{Boise State University, Boise, ID 83725}

\author{Peter P. Rohde}
\email[]{dr.rohde@gmail.com}
\homepage{http://www.peterrohde.org}
\affiliation{Centre for Engineered Quantum Systems, Department of Physics and Astronomy, Macquarie University, Sydney NSW 2113, Australia}
\affiliation{Centre for Quantum Computation and Intelligent Systems (QCIS), Faculty of Engineering \& Information Technology, University of Technology, Sydney, NSW 2007, Australia}

\date{\today}

\frenchspacing

\begin{abstract}
Quantum number-path entanglement is a resource for super-sensitive quantum metrology and in particular provides for sub-shotnoise or even Heisenberg-limited sensitivity. However, such number-path entanglement has thought to have been resource intensive to create in the first place --- typically requiring either very strong nonlinearities, or nondeterministic preparation schemes with feed-forward, which are difficult to implement. Very recently, arising from the study of quantum random walks with multi-photon walkers, as well as the study of the computational complexity of passive linear optical interferometers fed with single-photon inputs, it has been shown that such passive linear optical devices generate a superexponentially large amount of number-path entanglement. A logical question to ask is whether this entanglement may be exploited for quantum metrology. We answer that question here in the affirmative by showing that a simple, passive, linear-optical interferometer --- fed with only uncorrelated, single-photon inputs, coupled with simple, single-mode, disjoint photodetection --- is capable of significantly beating the shotnoise limit. Our result implies a pathway forward to practical quantum metrology with readily available technology.
\end{abstract}

\maketitle

Ever since the early work of Yurke \& Yuen it has been understood that quantum number-path entanglement is a resource for super-sensitive quantum metrology, allowing for sensors that beat the shotnoise limit \cite{bib:yurke1986input, bib:yuen1986generation}. Such devices would then have applications to super-sensitive gyroscopy \cite{bib:dowling1998correlated}, gravimetry \cite{bib:yurtsever2003interferometry}, optical coherence tomography \cite{bib:nasr2003demonstration}, ellipsometry \cite{bib:toussaint2004}, magnetometry \cite{bib:jones2009magnetic}, protein concentration measurements \cite{bib:crespi2012measuring}, and microscopy \cite{bib:rozema2014scalable, bib:israel2014supersensitive}. This line of work culminated in the analysis of the bosonic NOON state (\mbox{$(\ket{N,0}+\ket{0,N})/\sqrt{2}$}, where $N$ is the total number of photons), which was shown to be optimal for local phase estimation with a fixed, finite number of photons, and in fact allows one to hit the Heisenberg limit and the Quantum Cram{\'e}r-Rao Bound \cite{bib:holland1993interferometric, bib:lee2002quantum, bib:durkin2007local, bib:dowling2008quantum}.

Let us consider the NOON state as an example, where for this state in a two-mode interferometer we have the condition of all $N$ particles in the first mode (and none in the second mode) superimposed with all $N$ particles in the second mode (and none in the first mode). While such a state is known to be optimal for sensing, its generation is also known to be highly problematic and resource intensive. There are two routes to preparing high-NOON states: the first is to deploy very strong optical nonlinearities \cite{bib:gerry2001generation, bib:kapale2007bootstrapping}, and the second is to prepare them using measurement and feed-forward \cite{bib:lee2002linear, bib:vanmeter2007general, bib:cable2007efficient}. In many ways then NOON-state generators have had much in common with all-optical quantum computers and therefore are just as difficult to build \cite{bib:kok2007linear}. In addition to the complicated state preparation, typically a complicated measurement scheme, such as parity measurement at each output port, also had to be deployed \cite{bib:seshadreesan2013phase}. 

\begin{figure}[!htb]
\includegraphics[width=\columnwidth]{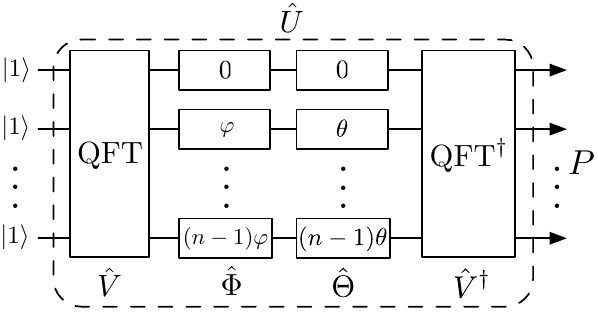}
\caption{Architecture of the quantum Fourier transform interferometer (QuFTI) for metrology using single-photon states. The input state comprises $n$ single photons, \mbox{$\ket{1}^{\otimes n}$}. The state evolves via the passive linear optics unitary \mbox{$\hat{U} = \hat{V} \cdot \hat{\Phi} \cdot \hat{\Theta} \cdot \hat{V}^\dag$}, where $\hat{V}$ is the quantum Fourier transform, $\hat\Phi$ is an unknown, linear phase gradient, and $\hat\Theta$ is a reference phase gradient used for calibration. At the output we perform a coincidence photodetection projecting on exactly one photon per output mode, measuring the observable $\hat{O}=(\ket{1}\bra{1})^{\otimes n}$, which, over many measurements, yields the probability distribution $P(\varphi)$ that acts as a witness for the unknown phase $\varphi$.} \label{fig:arch}
\end{figure}

Recently two independent lines of research, the study of quantum random walks with multi-photon walkers in passive linear-optical interferometers \cite{bib:mayer2011counting, bib:gard2013quantum, bib:gard2014inefficiency}, as well as the quantum complexity analysis of the mathematical sampling problem using such devices \cite{bib:AaronsonArkhipov10, bib:Chapter}, has led to a somewhat startling yet inescapable conclusion --- passive, multi-mode, linear-optical interferometers, fed with only uncorrelated single photon inputs in each mode (Fig. \ref{fig:arch}), produce quantum mechanical states of the photon field with path-number entanglement that grows superexponentially fast in the two resources of mode and photon-number \bibnote[See]{for another practical application of the complexity of linear optics, J. Huh, G. G. Guerreschi, B. Peropadre, J. R. McClean,
and A. Aspuru-Guzikl (2014), quant-ph/1412.8427.}. What is remarkable is that this large degree of number-path entanglement is not generated by strong optical nonlinearities, nor by complicated measurement and feed-forward schemes, but by the natural evolution of the single photons in the passive linear optical device. Whilst such devices are often described to have `non-interacting' photons in them, there is a type of photon-photon interaction generated by the demand of bosonic state symmetrization, which gives rise to the superexponentially large number-path entanglement via multiple applications of the Hong-Ou-Mandel effect \cite{bib:gard2014inefficiency}. It is known that linear optical evolution of single photons, followed by projective measurements, can give rise to `effective' strong optical nonlinearities, and we conjecture that there is indeed a hidden Kerr-like nonlinearity at work also in these interferometers \cite{bib:lapaire2003conditional}. Like boson-sampling \cite{bib:AaronsonArkhipov10}, and unlike universal quantum computing schemes such as that by Knill, Laflamme, and Milburn \cite{bib:LOQC}, this protocol is deterministic and does not require any ancillary photons.

The advantage of such a setup for quantum metrology is that resources for generating and detecting single photons have become quite standardized and relatively straightforward to implement in the lab \cite{bib:matthews2011heralding, bib:spring2013boson, bib:Broome2012, bib:crespi2013integrated, bib:ralph2013quantum, bib:motes2013spontaneous, bib:spagnolo2014experimental}. The community then is moving towards single photons, linear interferometers, and single-photon detectors all on a single, integrated, photonic chip, which then facilitates a roadmap for scaling up devices to large numbers of modes and photons. If all of this work could be put to use for quantum metrology, then a road to scalable metrology with number states would be at hand. 

It then becomes a natural question to ask --- since number-path entanglement is known to be a resource for quantum metrology --- can a passive, multi-mode interferometer, fed only with easy-to-generate uncorrelated single photons in each mode, followed by uncorrelated single-photon measurements at each output, be constructed to exploit this number-path entanglement for super-sensitive (sub-shotnoise) operation? The answer is indeed yes, as we shall now show.

The phase-sensitivity, $\Delta\varphi$, of a metrology device can be defined in terms of the standard error propagation formula as, 
\begin{eqnarray} \label{eq:phaseSensitivity}
\Delta\varphi = \frac{\sqrt{\expec{\hat{O}^2}-\expec{\hat{O}}^2}}{\left|\frac{\partial\expec{\hat{O}}}{\partial\varphi}\right|}, 
\end{eqnarray}
where $\expec{\hat{O}}$ is the expectation of the observable being measured and $\varphi$ is the unknown phase we seek to estimate.

The photons evolve through a unitary network according to $U a_i^{\dag} U^{\dag} = \sum_j U_{ij} a_j^{\dag}$. In our protocol, we construct the $n$-mode interferometer $\hat{U}$ to be,
\begin{equation} \label{eq:U}
\hat{U} = \hat{V} \cdot \hat{\Phi} \cdot \hat{\Theta} \cdot \hat{V}^{\dag},
\end{equation}
which we call the quantum fourier transform interferometer (QuFTI) because $\hat{V}$ is the $n$-mode quantum Fourier transform matrix, with matrix elements given by,
\begin{equation}
\mathrm{V}_{j,k}^{(n)} = \frac{1}{\sqrt{n}}\mathrm{exp}\left[\frac{- 2 i j k \pi}{n}\right].
\end{equation} 
$\hat\Phi$ and $\hat\Theta$ are both diagonal matrices with linearly increasing phases along the diagonal represented by,
\begin{eqnarray} \label{eq:PhiTheta}
\Phi_{j,k} = \delta_{j,k} \exp{\Big[i(j-1)\varphi\Big]} \nonumber \\
\Theta_{j,k} = \delta_{j,k} \exp{\Big[i(j-1)\theta\Big]},
\end{eqnarray}
where $\varphi$ is the unknown phase one would like to measure and $\theta$ is the control phase. $\hat\Theta$ is introduced as a reference, which can calibrate the device by tuning $\theta$ appropriately. To see this tuning we combine $\hat\Phi$ and $\hat\Theta$ into a single diagonal matrix with a gradient given by,
\begin{equation}
\Phi_{j,k}\cdot\Theta_{j,k} = \delta_{j,k} \exp{\bigg[i(j-1)(\varphi+\theta)\bigg]}.
\end{equation}
The control phase $\theta$ can shift this gradient to the optimal measurement regime, which can be found by minimizing $\Delta\varphi$ with respect to $n$ and $\varphi$. Since this is a shift according to a known phase, we can for simplicity assume (and without loss of generality) that $\varphi$ is in the optimal regime for measurements and $\theta=0$. Thus, $\hat\Theta=\hat{I}$ and is left out of our analysis for simplicity.

In order to understand how such a linearly increasing array of unknown phase shifts may be arranged in a practical device, it is useful to consider a specific example. Let us suppose that we are to use the QuFTI as an optical magnetometer. We consider an interferometric magnetometer of the type discussed in \cite{bib:scully1992high} where each of the sensing modes of the QuFTI contains a gas cell of Rubidium prepared in a state of electromagnetically induced transparency whereby a photon passing through the cell at the point of zero absorption in the electromagnetically induced transparency spectrum acquires a phase shift   that is proportional to the product of an applied uniform (but unknown) magnetic field   and the length of the cell. We assume that the field is uniform across the QuFTI, as would be the case if the entire interferometer was constructed on an all optical chip and the field gradient across the chip were negligible. Since we are carrying out local phase measurements (not global) we are not interested in the magnitude of the magnetic field but wish to know if the field changes and if so by how much. (Often we are interested in if the field is oscillating and with what frequency.) Neglecting other sources of noise then in an ordinary Mach-Zehnder interferometer this limit would be set by the photon shotnoise limit. To construct a QuFTI with the linear cascade of phase shifters, as shown in Fig. \ref{fig:arch}, we simply increase the length of the cell by integer amounts in each mode. The first cell has length $L$, the second length $2L$, and so forth. This will then give us the linearly increasing configuration of unknown phase shifts required for the QuFTI to beat the SNL. 

One might question why one would employ a phase gradient rather than just a single phase. Investigation into using a single phase in $\hat\Phi$ indicates that this yields no benefit. We conjecture that this is because the number of paths interrogating a phase in a single mode is not superexponential as is the case when a phase gradient is employed.

The interferometer may always be constructed efficiently following the protocol of Reck \emph{et al.} \cite{bib:Reck94}, who showed that an \mbox{$n\times n$} linear optics interferometer may be constructed from $O(n^2)$ linear optical elements (beamsplitters and phase-shifters), and the algorithm for determining the circuit has runtime polynomial in $n$. Thus, an experimental implementation of our protocol may always be efficiently realized.

The input state to the device is \mbox{$\ket{1}^{\otimes n}$}, i.e. single photons inputed in each mode. If \mbox{$\varphi=0$} then $\hat\Phi=\hat{I}$ and thus $\hat{U}=\hat{V}\cdot\hat{I}\cdot\hat{V}^{\dag}=\hat{I}$. In this instance, the output state is exactly equal to the input state, \mbox{$\ket{1}^{\otimes n}$}. Thus, if we define $P$ as the coincidence probability of measuring one photon in each mode at the output, then \mbox{$P=1$} when \mbox{$\varphi=0$}. When \mbox{$\varphi\neq 0$}, in general \mbox{$P<1$}. Thus, intuitively, we anticipate that \mbox{$P(\varphi)$} will act as a witness for $\varphi$. 

In the protocol, assuming a lossless device, no measurement events are discarded. Upon repeating the protocol many times, let $x$ be the number of measurement outcomes with exactly one photon per mode, and $y$ be the number of measurement outcomes without exactly one photon per mode. Then $P$ is calculated as $P=x/(x+y)$. Thus, all measurement outcomes contribute to the signal and none are discarded. Note that, due to preservation of photon-number and the fact that we are considering the anti-bunched outcome, $P(\varphi)$ may be experimentally determined using non-number-resolving detectors if the device is lossless.  If the device is assumed to be lossy, then number-resolving detectors would be necessary to distinguish between an error outcome and one in which more than one photon exits the same mode.  The circuit for the architecture is shown in Fig. \ref{fig:arch}.

The state at the output to the device is a highly path-entangled superposition of $\binom{2n-1}{n}$ terms, which grows superexponentially with $n$.  This corresponds to the number of ways to add $n$ non-negative integers whose sum is $n$, or equivalently, the number of ways to put $n$ indistinguishable balls into $n$ distinguishable boxes. We conjecture that this superexponential path-entanglement yields improved phase-sensitivity as the paths query the phases a superexponential number of times.

The observable being measured is the projection onto the state with exactly one photon per output mode, \mbox{$\hat{O} = (\ket{1}\bra{1})^{\otimes n}$}. Thus, \mbox{$\langle\hat{O}\rangle = \langle\hat{O}^2\rangle = P$}. And, the phase-sensitivity estimator reduces to,
\begin{equation} \label{eq:phaseSenP}
\Delta\varphi = \frac{\sqrt{P - P^2}}{\left|\frac{\partial P}{\partial \varphi}\right|}.
\end{equation}

Following the result of \cite{bib:ScheelPerm}, $P$ is related to the permanent of $\hat U$ as,
\begin{equation}
P = \big|\mathrm{Per}(U)\big|^2.
\end{equation}
Here the permanent of the full \mbox{$n\times n$} matrix is computed, since exactly one photon is going into and out of every mode. This is unlike the boson-sampling protocol \cite{bib:AaronsonArkhipov10} where permanents of sub-matrices are computed.

We will now  examine the structure of this permanent. The matrix form for the $n$-mode unitary $\hat U^{(n)}$ is given by,
\begin{equation} \label{eq:Ujk}
U_{j,k}^{(n)} =\frac{1-e^{i n\varphi}}{n\left(e^{\frac{2 i \pi(j-k)}{n}}-e^{i \varphi}\right)},
\end{equation}
as derived in App. \ref{app:Ujk}. Taking the permanent of this matrix is challenging as calculating permanents are in general \mbox{\textbf{\#P}-hard}. However, based on calculating $\mathrm{Per}(\hat U^{(n)})$ for small $n$, we observe the empirical pattern,
\begin{equation} \label{eq:permU}
\mathrm{Per}(\hat{U}^{(n)})= \frac{1}{n^{n-1}}\prod_{j=1}^{n-1}\Big[je^{i n \varphi}+n-j\Big],
\end{equation}
as conjectured in App. \ref{app:series}. This analytic pattern we observe is not a proof of the permanent, but an empirical pattern --- a conjecture --- that has been verified by brute force to be correct up to $n=25$. Although we don't have a proof beyond that point, $n=25$ is well beyond what will be experimentally viable in the near future, and thus the pattern we observe is sufficient for experimentally enabling super-sensitive metrology with technology available in the foreseeable future.

\begin{figure}[!htb]
\includegraphics[width=\columnwidth]{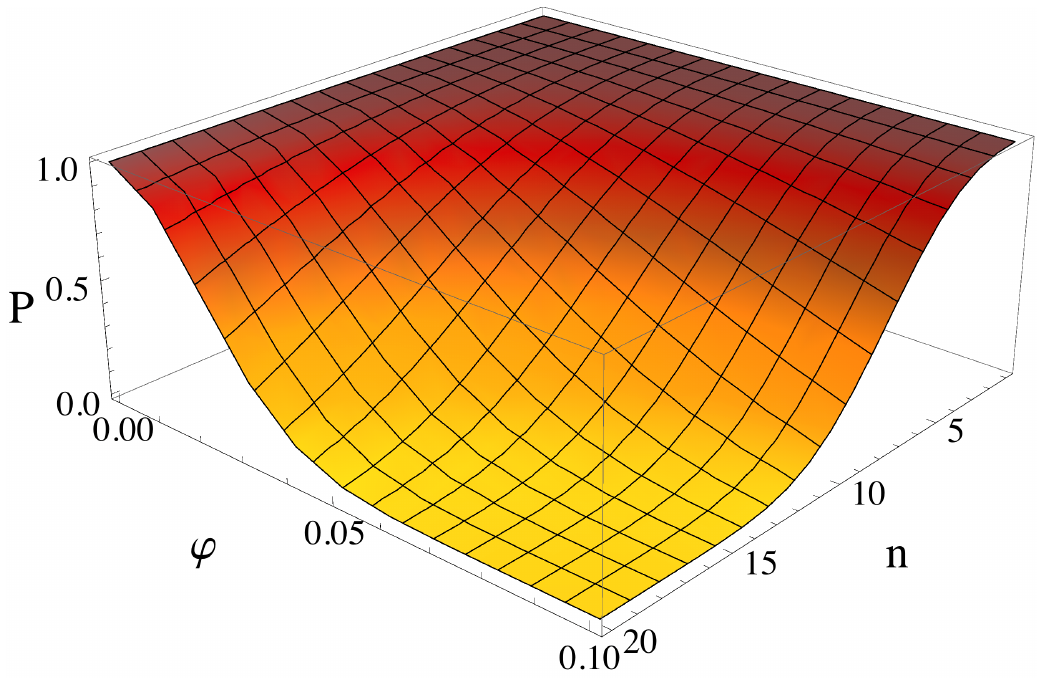}
\caption{Coincidence photodetection probability $P$ against the unknown phase $\varphi$ and the number of photons and modes $n$. As $n$ increases, the dependence of $P$ on $\varphi$ increases, resulting in improved phase-sensitivity.} \label{fig:P}
\end{figure}

Following as a corollary to the previous conjecture, the coincidence probability of measuring one photon in each mode is,
\begin{eqnarray} \label{eq:P_Result}
P &=& \Big|\mathrm{Per}(\hat{U}^{(n)})\Big|^2 \nonumber \\
&=& \frac{1}{n^{2n-2}}\prod_{j=1}^{n-1} \Big[a_n(j)\mathrm{cos}(n\varphi)+b_n(j) \Big],
\end{eqnarray} 
as shown in App. \ref{app:P}, where
\begin{eqnarray}
a_n(j) &=& 2j(n-j), \nonumber \\
b_n(j) &=& n^2-2jn+2j^2.
\end{eqnarray}
The dependence of $P$ on $n$ and $\varphi$ is shown in Fig. \ref{fig:P}.

It then follows that,
\begin{equation} \label{eq:dP}
\left|\frac{\partial P}{\partial \varphi}\right| = nP\big|\mathrm{sin}(n\varphi)\big|\sum_{j=1}^{n-1} \left|\frac{a_n(j)}{a_n(j)\mathrm{cos}(n\varphi)+b_n(j)}\right|,
\end{equation}
as shown in App. \ref{app:dP}.

\begin{figure}[!htb] 
\includegraphics[width=\columnwidth]{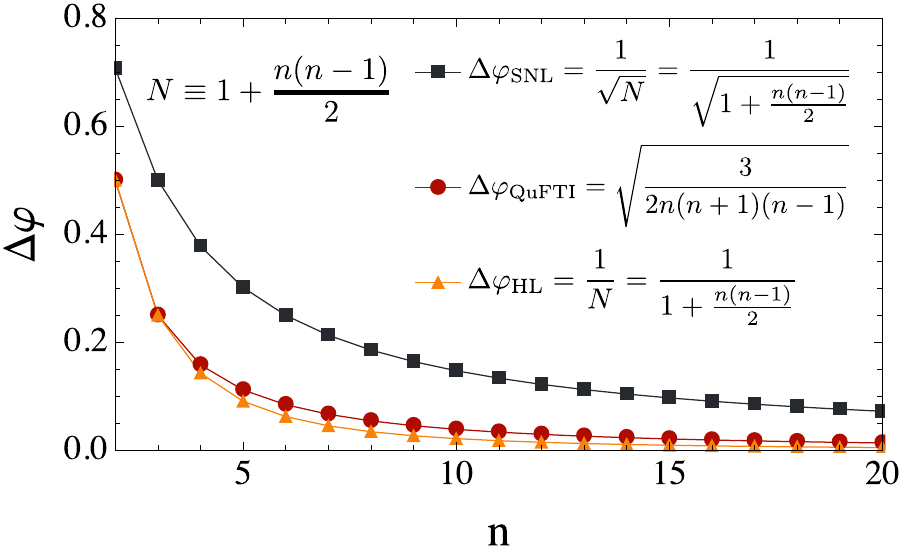}
\caption{Phase-sensitivity $\Delta\varphi$ against the number of photons $n$ (red circles). The shotnoise limit of $1/\sqrt{N}$ (black squares) and Heisenberg limit of $1/N$ (orange triangles) are shown for comparison. The QuFTI exhibits phase-sensitivity significantly better than the shotnoise limit, and only slightly worse than the Heisenberg limit. \label{fig:Delta}}  
\end{figure}

Finally, we wish to establish the scaling of $\Delta\varphi$. With a small $\varphi$ approximation (\mbox{$\mathrm{sin}(\varphi)\approx\varphi$}, \mbox{$\mathrm{cos}(\varphi)\approx 1-\frac{1}{2}\varphi^2$}) we find,
\begin{eqnarray} \label{eq:DeltaVarPhi}
\Delta\varphi &=& \sqrt{\frac{3}{2n(n+1)(n-1)}} \\ \nonumber
&=& \frac{1}{2\sqrt{{{n+1}\choose{3}}}},
\end{eqnarray}
as shown in App. \ref{app:dphi}. Thus, the phase sensitivity scales as \mbox{$\Delta\varphi = O(1/n^{3/2})$} as shown in Fig. \ref{fig:Delta}.

We would like to compare the performance of our QuFTI to an equivalent multimode interferometer baseline for which we will construct the shotnoise limit (SNL) and Heisenberg limit (HL). This is a subtle comparison, due to the linearly increasing unknown phase-shifts, \{\mbox{$0,\varphi,\dots,(n-1)\varphi$}\}, that the QuFTI requires to operate. The mathematical relation is shown in Fig. \ref{fig:Delta}, where we have converted the number of resources, $N$, to the number of photons, $n$. There is disagreement on how such resources should be counted. This is the method, which we call Ordinal Resource Counting (ORC), that we feel most fairly counts our resources. A more detailed supporting discussion can be found in App. \ref{app:counting}.

While computing the sensitivity using the standard error propagation formula of Eq. \ref{eq:phaseSensitivity} provides clear evidence that our scheme does indeed beat the SNL, it would be instructive to carry out a calculation of the quantum Fisher information and thereby provide the quantum Cram{\'e}r-Rao bound, which would be a true measure of the best performance of this scheme possible, according to the laws of quantum theory. However, due to the need to compute the permanent of large matrices with complex entries, this calculation currently remains intractable. We will continue to investigate such a computation for a future work. In general, analytic solutions to matrix permanents are not possible. In this instance, the analytic result is facilitated by the specific structure of the QuFTI unitary. Other inhomogeneous phase gradients may yield analytic results, but we leave this for future work.

In App. \ref{app:efficiency} we discuss the efficiency of the QuFTI protocol and in App. \ref{app:dephasing} we analyse dephasing, which is a source of decoherence, and find that the QuFTI protocol is far more robust against dephasing than the NOON state is.

We have shown that a passive linear optics network fed with single-photon Fock states may implement quantum metrology with phase-sensitivity that beats the shotnoise limit. Unlike other schemes that employ exotic states such as NOON states, which are notoriously difficult to prepare, single-photon states may be readily prepared in the laboratory using present-day technology. This new approach to metrology via easy-to-prepare single-photon states and disjoint photodetection provides a road towards improved quantum metrology with frugal physical resources.

\begin{acknowledgments}
We thank Dominic Berry, Alexei Gilchrist, \& Michael Bremner for helpful discussions.  We would also like to thank Kaushik Seshadreesan and Hwang Lee for helpful discussion on resource counting.  KRM and PPR would like to acknowledge the Australian Research Council Centre of Excellence for Engineered Quantum Systems (Project number CE110001013). P. P. R acknowledges support from Lockheed Martin. JPO, EJR, and JPD would like to acknowledge support from the Air Force Office of Scientific research, the Army Research Office, and the National Science Foundation.
\end{acknowledgments}

\appendix

\section{Proof of $U_{j,k}^{(n)}$} \label{app:Ujk}

Beginning from Eq. \ref{eq:U} and setting $\hat{\Theta}=\hat{I}$,
\begin{eqnarray}
U_{j,k}^{(n)} &=& (\hat{V}\hat{\Phi}\hat{V^{\dag}})_{j,k} \nonumber \\ 
&=& \sum_{l,m=1}^{n}V_{j,l}\Phi_{l,m}V_{m,k}^{\dag} \nonumber \\
&=& \sum_{l,m=1}^{n} \underbrace{\frac{e^{- 2 i j l \pi/n}}{\sqrt{n}}}_{V_{j,l}}\underbrace{\delta_{l,m}e^{i(l-1)\varphi}}_{\Phi_{l,m}}\underbrace{\frac{e^{2 i m k \pi/n}}{\sqrt{n}}}_{V_{m,k}^{\dag}} \nonumber \\
&=& \frac{1}{n} \sum_{l=1}^{n} e^{\frac{- 2 i j l \pi}{n}} e^{i (l-1) \varphi} e^{\frac{2 i l k \pi}{n}}\nonumber \\
&=& \frac{1}{n} \sum_{l=1}^{n} e^{\frac{2 i l(k-j) \pi}{n} + i(l-1)\varphi} \nonumber \\
&=& e^{\frac{2 i (k-j) \pi}{n}}\frac{1}{n} \sum_{l=0}^{n-1} (e^{\frac{2 i (k-j) \pi}{n} + i\varphi})^l. \nonumber 
\end{eqnarray}
From the geometric series, it follows,
\begin{eqnarray}
U_{j,k}^{(n)}&=& \frac{1}{n(e^{\frac{2 i (j-k) \pi}{n}})}\frac{1-e^{i n\varphi}}{\left(1-e^{\frac{2 i (k-j) \pi}{n} +i \varphi}\right)}, \label{eq:U} \nonumber \\
&=& \frac{1-e^{i n\varphi}}{n\left(e^{\frac{2 i \pi(j-k)}{n}}-e^{i \varphi}\right)}
\end{eqnarray}
which is what we set out to prove.
which is Eq. \ref{eq:Ujk} that we set out to prove, where the last line follows from the geometric series.

\section{Conjecture for the Analytic Form of Per($\hat U^{(n)}$)} \label{app:series}

Our goal is to find the analytic form for Per($\hat U^{(n)}$) where $U_{j,k}^{(n)}$ is as in Eq. \ref{eq:U}.  We can perform a brute force calculation to obtain the analytic form for small $n$.  Doing so up to $n=6$ yields:
\begin{table}[H]
\begin{tabular}{|c|c|}
\hline
 $n$ & Per$(\hat U^{(n)})$ \nonumber \\
 \hline
 1 & 1 \nonumber \\
 2 & $e^{i \phi } \cos (\phi )$ \\
 3 & $\ \frac{1}{9} \left(2+e^{3 i \phi   }\right) \left(1+2 e^{3 i \phi   }\right)$ \\
 4 & $\frac{1}{32} \left(1+e^{4 i \phi   }\right) \left(3+e^{4 i \phi }\right)    \left(1+3 e^{4 i \phi }\right)$ \\
 5 & $\frac{1}{625} \left(4+e^{5 i \phi
   }\right) \left(3+2 e^{5 i \phi   }\right) \left(2+3 e^{5 i \phi   }\right) \left(1+4 e^{5 i \phi   }\right)$ \\
 6 & $\frac{1}{648} \left(1+e^{6 i \phi    }\right) \left(2+e^{6 i \phi }\right)    \left(5+e^{6 i \phi }\right)    \left(1+2 e^{6 i \phi }\right)    \left(1+5 e^{6 i \phi }\right)$ 
 \\
 \hline
\end{tabular}
\end{table} 
One can see the pattern that emerges is of the form:
\begin{equation} \label{eq:permU2} 
\mathrm{Per}(\hat{U}^{(n)})= \frac{1}{n^{n-1}}\prod_{j=1}^{n-1}\Big[je^{i n \varphi}+n-j\Big],
\end{equation}
which is Eq. \ref{eq:permU} that we set out to show. This equation has been verified analytically up to $n=16$ and up to $n=25$ numerically..

\section{Calculation of $P$} \label{app:P}
Assuming our conjecture in Eq. \ref{eq:permU} holds, we can compute the coincidence probability of measuring one photon in each mode at the output,
\begin{eqnarray}
P &=& \big|\mathrm{Perm}(U^{(n)})\big|^2 \nonumber \\
&=& \left|\frac{1}{n^{n-1}} \prod_{j=1}^{n-1}\left(je^{i n \varphi}+n-j\right)\right|^2 \nonumber \\
&=& \frac{1}{n^{2n-2}} \prod_{j=1}^{n-1}\Big|\left(je^{i n \varphi}+n-j\right)\Big|^2 \nonumber \\
&=& \frac{1}{n^{2n-2}} \prod_{j=1}^{n-1}\Big|j\mathrm{cos}(n\varphi)+ij\mathrm{sin}(n\varphi)+n-j\Big|^2 \nonumber \\
&=& \frac{1}{n^{2n-2}} \prod_{j=1}^{n-1}\Big|\underbrace{j\mathrm{cos}(n\varphi)+(n-j)}_\mathrm{Re}+i\underbrace{j\mathrm{sin}(n\varphi)}_\mathrm{Im}\Big|^2. \nonumber \\
\end{eqnarray} 
Invoking the property that \mbox{$|z|^2= \mathrm{Re}(z)^2+\mathrm{Im}(z)^2$}, where \mbox{$z\in \mathbb{C}$}, 
\begin{eqnarray} \label{eq:Pproof}
P &=& \frac{1}{n^{2n-2}} \prod_{j=1}^{n-1} \Big[\big(j\mathrm{cos}(n\varphi)+(n-j)\big)^2+j^2\mathrm{sin}^2(n\varphi)\Big] \nonumber \\
&=& \frac{1}{n^{2n-2}} \prod_{j=1}^{n-1} \Big[\underbrace{j^2\mathrm{cos}^2(n\varphi)+j^2\mathrm{sin}^2(n\varphi)}_{=j^2} \nonumber \\
&+& 2j(n-j)\mathrm{cos}(n\varphi)+(n-j)^2 \Big] \nonumber \\
&=& \frac{1}{n^{2n-2}} \prod_{j=1}^{n-1} \Big[j^2 + 2j(n-j)\mathrm{cos}(n\varphi)+(n-j)^2 \Big]\nonumber \\
&=& \frac{1}{n^{2n-2}} \prod_{j=1}^{n-1} \Big[\underbrace{2j(n-j)}_{a_n(j)}\mathrm{cos}(n\varphi)+\underbrace{n^2-2jn+2j^2}_{b_n(j)} \Big] \nonumber \\
&=& \frac{1}{n^{2n-2}} \prod_{j=1}^{n-1} \Big[a_n(j)\mathrm{cos}(n\varphi)+b_n(j) \Big], \nonumber \\
\end{eqnarray} 
which is Eq. \ref{eq:P_Result} that we set out to show.

\section{Calculation of $\left|\frac{\partial P}{\partial \varphi}\right|$} \label{app:dP}

From Eq. \ref{eq:Pproof}, exploiting the logarithm product rule,
\begin{eqnarray}
\mathrm{ln}(P) &=& \underbrace{\mathrm{ln}\left(\frac{1}{n^{2n-2}}\right)}_{C} + \mathrm{ln}\left(\prod_{j=1}^{n-1} \Big[a_n(j)\mathrm{cos}(n\varphi) + b_n(j) \Big] \right) \nonumber \\
&=& C + \sum_{j=1}^{n-1} \mathrm{ln}\Big[a_n(j)\mathrm{cos}(n\varphi)+b_n(j) \Big],
\end{eqnarray} 
where $C$ is a constant. Now the derivative becomes, 
\begin{eqnarray}
\frac{1}{P}\frac{\partial P}{\partial \varphi} &=& -\sum_{j=1}^{n-1} \frac{na_n(j)\mathrm{sin}(n\varphi)}{a_n(j)\mathrm{cos}(n\varphi)+b_n(j)} \nonumber \\
\frac{\partial P}{\partial \varphi} &=& -nP\mathrm{sin}(n\varphi)\sum_{j=1}^{n-1} \frac{a_n(j)}{a_n(j)\mathrm{cos}(n\varphi)+b_n(j)}.\nonumber \\
\end{eqnarray} 
Thus,
\begin{equation}
\left|\frac{\partial P}{\partial\varphi}\right|=nP\big|\mathrm{sin}(n\varphi)\big|\sum_{j=1}^{n-1} \left|\frac{a_n(j)}{a_n(j)\mathrm{cos}(n\varphi)+b_n(j)}\right|,
\end{equation}
which is Eq. \ref{eq:dP} that we set out to show.

\section{Calculation of $\Delta\varphi$ in the small angle approximation} \label{app:dphi}
We wish to compute $\Delta\varphi$ in the limit that $n\varphi\ll1$. Then $P$ in the small angle regime of Eq. \ref{eq:P_Result} becomes,
\begin{eqnarray}
\label{eq:papprox}
P&\approx& \frac{1}{n^{2n-2}}\prod_{j=1}^{n-1} \bigg[a_n(j)\Big(1-\frac{1}{2}(n\varphi)^2\Big)+b_n(j)\bigg] \nonumber \\ 
&=& \frac{1}{n^{2n-2}}\prod_{j=1}^{n-1}\bigg[ n^2-(nj+j^2)n^2\varphi^2 \bigg]\nonumber \\ 
&=& \prod_{j=1}^{n-1}\Big[1-(nj+j^2)\varphi^2\Big],
\end{eqnarray}
where $\cos(n\varphi)$ is expanded to the first nonconstant term in its Taylor series. This product has the form of a binomial expansion. Dropping terms above order $\varphi^2$, $P$ reduces to,
\begin{eqnarray}
\label{eq:pfinal}
P&\approx& 1-\varphi^2\sum_{j=1}^{n-1}\Big[nj+j^2\Big] \nonumber \\
&=& 1-\varphi^2\Big[\frac{1}{6}(n-1)n(n+1)\Big] \nonumber \\
&=&1-k(n)\varphi^2,
\end{eqnarray}
where $k(n)=\frac{1}{6}n(n-1)(n+1)\geq0$ $\forall$ $n\geq1$.  From Eq. \ref{eq:pfinal} we can easily compute $P^2$ and $\big|\frac{\partial P}{\partial \varphi}\big|$ to be,
\begin{eqnarray}
P^2&\approx& 1-2k(n)\varphi^2 \\
\left|\frac{\partial P}{\partial \varphi}\right|&=&2k(n)|\varphi|,
\end{eqnarray}
where we have again dropped terms above order $\varphi^2$. Using Eq. \ref{eq:phaseSenP} the phase sensitivity $\Delta\varphi$ in the small angle regime is,
\begin{eqnarray}
\Delta\varphi &=& \frac{\sqrt{P-P^2}}{\left|\frac{\partial P}{\partial\varphi}\right|} \nonumber \\
&=&\frac{\sqrt{\Big(1-k(n)\varphi^2\Big)-\Big(1-2k(n)\varphi^2\Big)}}{2k(n)|\varphi|} \nonumber \\
&=&\frac{\sqrt{k(n)\varphi^2}}{2k(n)|\varphi|} \nonumber \\
&=& \frac{1}{2\sqrt{k(n)}} \nonumber \\
&=&\sqrt{\frac{3}{2{(n-1)n(n+1)}}},
\end{eqnarray}
which is Eq. \ref{eq:DeltaVarPhi} that we set out to show.

\section{Discussion of Ordinal Resource Counting (ORC)} \label{app:counting}

We would like to compare the performance of our QuFTI to an equivalent multimode interferometer baseline for which we will construct the shotnoise limit (SNL) and Heisenberg limit (HL). This is a subtle comparison, due to the linearly increasing unknown phase-shifts, \{\mbox{$0,\varphi,\dots,(n-1)\varphi$}\}, that the QuFTI requires to operate. There is a long and muddled history of increasing the interrogation time (or here length) of the probe particles with the unknown phase-shift followed by an incorrect reckoning of the true resources. Here we shall use a protocol we call Ordinal Resource Counting (ORC) whereby all resources, such as number of `calls' to the phase-shifter $\varphi$, are converted to the `currency' of the resource that is most precious to us, namely photon-number. We do this as follows. 

\begin{figure}[!htb]
\includegraphics[width=0.8\columnwidth]{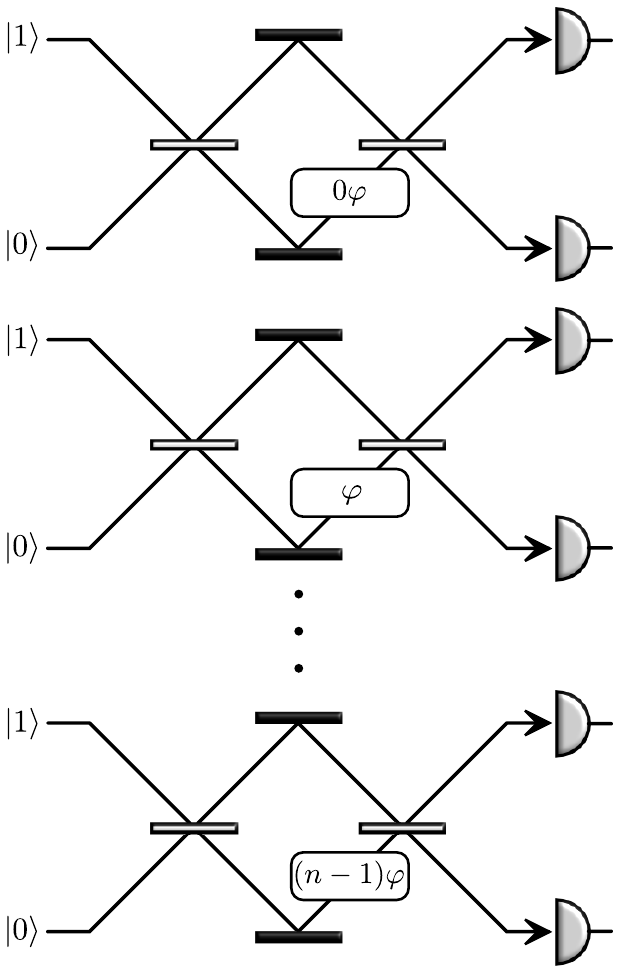}
\caption{$n$ instances of two-mode Mach-Zehnder interferometers, with a linearly increasing phase gradient. This system has the same configuration of phases as the QuFTI, but the photons are not allowed to interfere, and thus has minimal number-path entanglement.} \label{fig:resources1}
\end{figure}

First we must construct a multimode interferometer with $n$ photon inputs that provides the baseline if the photons remain uncorrelated and the number-path entanglement remains minimal. Such a comparator is shown in Fig. \ref{fig:resources1}, and consists of $n$, two-mode Mach-Zehnder Interferometers (MZI) in a vertical cascade, fed with single-photon inputs, with the same linearly increasing unknown phase-shift sequence as the QuFTI. Since the MZIs are disconnected, the number-path entanglement remains constant and minimal, and of the form \mbox{$(\ket{1,0}+\ket{0,1})/\sqrt{2}$} inside each MZI. 

\begin{figure}[!htb]
\includegraphics[width=0.9\columnwidth]{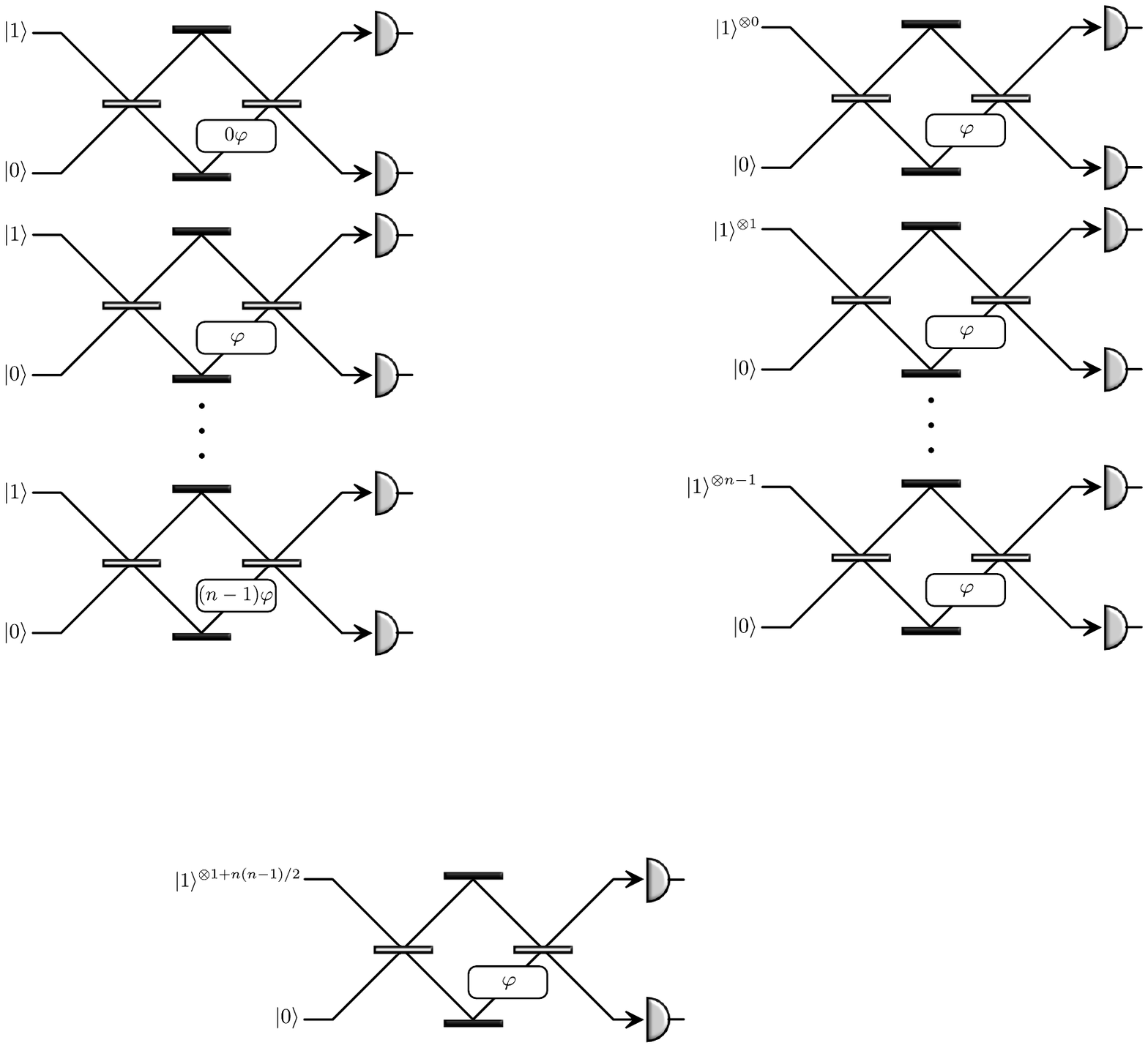}
\caption{Noting that a single photon interrogating a phase-shift of $n\varphi$ is equivalent to $n$ independent interrogations of $\varphi$, Fig. \ref{fig:resources1} can be represented in terms of the resource of photons as shown here. Here $\ket{1}^{\otimes j}$ means that $j$ independent (i.e distinguishable) photons have been prepared.} \label{fig:resources2}
\end{figure}

Now to convert the linearly increasing interrogation lengths of the unknown phase-shifts, we note that a single photon interrogating a phase-shift of say $2\varphi$ is equivalent to a single photon interrogating a single phase-shift $\varphi$ twice, which is in turn equivalent to two uncorrelated photons entering the same port of the MZI containing a single phase-shift of $\varphi$. In this way we may convert `number of interrogations of the phase-shifter' into the currency of `number of photons' to carry out a fair reckoning of the resources. Following this logic we are led to Fig. \ref{fig:resources2} showing a cascade of MZIs where the linearly increasing phase-shifters are replaced with a single phase-shifter of $\varphi$ and the single photons at the MZI inputs are replaced with a linearly increasing number of photons. Then the `number of interrogations of the phase-shifter' becomes $n(n-1)/2$, but there is an additional photon that is part of the QuFTI resources so our total number of resources becomes,
\begin{equation} \label{eq:N}
N\equiv1+\frac{n(n-1)}{2}.
\end{equation}

\begin{figure}[!htb]
\includegraphics[width=0.9\columnwidth]{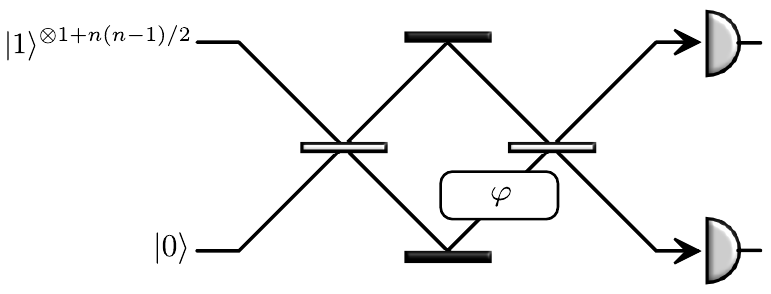}
\caption{Grouping all the independent interferometers in Fig. \ref{fig:resources2} together and including the extra photon from the QuFTI model, we obtain a single MZI with \mbox{$1+n(n-1)/2$} independent photons as input. This configuration achieves the shotnoise limit, and thus provides a benchmark for comparing our QuFTI protocol against the shotnoise and Heisenberg limits, with photons as the resource being counted.} \label{fig:resources3}
\end{figure}

Next we note that this cascade of $n$ MZIs in Fig. \ref{fig:resources2} may be replaced with a single MZI, shown in Fig. \ref{fig:resources3}, where the input is now an ordinal grouped ranking of the uncorrelated photons following the same pattern as in Fig. \ref{fig:resources2}. Hence in the configuration in Fig. \ref{fig:resources3} we have a single MZI with vacuum entering the lower port, a stream of $N$ uncorrelated photons entering the upper port, and a single phase-shifter $\varphi$ between the beamsplitters. It is well-known that for this configuration the sensitivity of this system scales as the SNL \cite{bib:scully1993quantum, bib:dowling1998correlated}, namely,
\begin{equation}
\Delta\varphi_\mathrm{SNL} = \frac{1}{\sqrt{N}} = \frac{1}{\sqrt{1+\frac{n(n-1)}{2}}}.
\end{equation} 
This then provides us a fair reckoning of the SNL to be used gauging the performance of the QuFTI.

Finally, if instead we were to maximally path-number entangle these resources into a NOON state of the form \mbox{$(\ket{N,0}+\ket{0,N})/\sqrt{2}$} (just to the right of the first beam splitter but before the phase-shifter) the sensitivity then becomes Heisenberg limited,
\begin{equation}
\Delta\varphi_\mathrm{HL} = \frac{1}{N} = \frac{1}{1+\frac{n(n-1)}{2}},
\end{equation}
which is a sensitivity known to saturate the Quantum Cram{\'e}r-Rao Bound (CRB) for sensitivity in local phase estimation with $N$ photons \cite{bib:lee2002quantum, bib:durkin2007local}. As the CRB is the best one may do, according to the laws of quantum mechanics, then in this case the HL is optimal. As discussed, the performance of the QuFTI falls between the SNL and the HL, but with the feature of not having to do anything resource intensive such as preparing a high-NOON state. 

Thus the SNL and the HL, computed via this Ordinal Resource Counting method, provides the fairest comparison of sensitivity performance of the QuFTI with such ambiguities such as how to handle `number of calls to the phase-shifter' removed by replacing such a notion with `number of photons' inputted into the interferometer. 

\section{Efficiency} \label{app:efficiency}

In the presence of inefficient photon sources and photo-detectors the success probability of the protocol will drop exponentially with the number of photons. Specifically, if $\eta_s$ and $\eta_d$ are the source and detection efficiencies respectively, the success probability of the protocol is $\eta = (\eta_s \eta_d)^n$. Current cutting edge transition edge detectors operate at 98\% efficiency, with negligible dark count \cite{bib:fukuda2011titanium}. SPDC sources are the standard photon-source technology but they are non-deterministic. However, there are techniques that can greatly improve the heralding efficiency up to 42\% at 2.1 MHz \cite{bib:LPOR201400404}. Also, other source technologies, such as quantum dot sources are becoming viable with efficiencies also up to 42\% \cite{bib:Maier14}. For $n=10$, which is already well beyond current experiments, this yields $\eta= (0.98*0.42)^{10} \approx 0.00014$, which is about 300 successful experimental runs per second when operating with 2.1 MHz sources. 

\section{Dephasing} \label{app:dephasing}

A form of decoherence to consider is dephasing. Dephasing in our work may be modelled with the result of Bardhan \emph{et al.} \cite{bib:Bardhan2013}, whereby dephasing occurs on each mode separately. When considering our example of a magnetometer, dephasing would occur in the magnetic field cells where atomic fluctuations may occur that differ between cells.  In the rest of the interferometer, dephasing can be made very close to zero, particularly on an all optical chip. 

\begin{figure}[t]
\includegraphics[width=0.9\columnwidth]{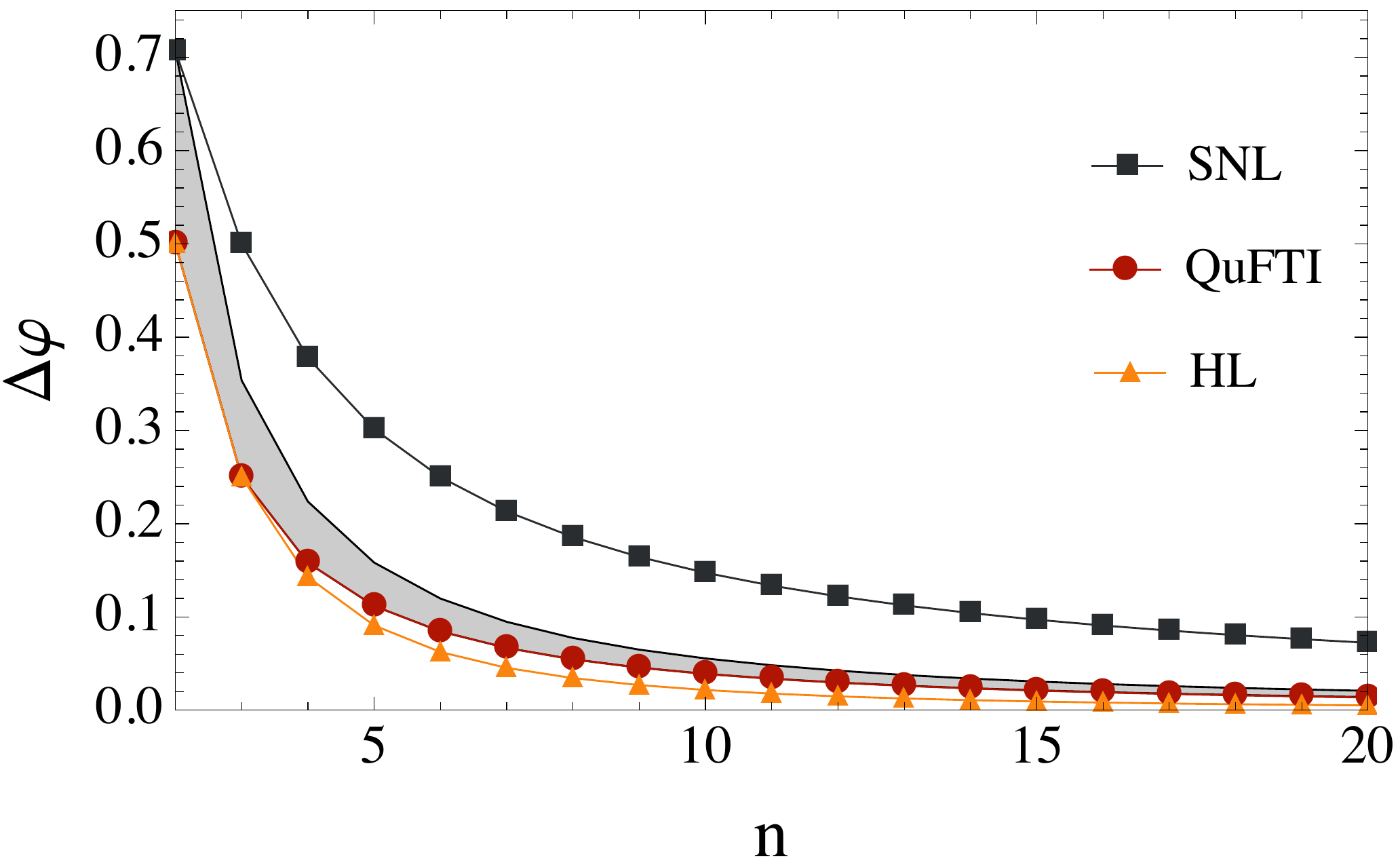}
\caption{Dephasing for $\varphi=0.01$.  The shaded region represents the phase sensitivity for the QuFTI where $0\leq\chi\leq 0.01$.} \label{fig:dephasing}
\end{figure}

\begin{figure}[b]
\includegraphics[width=0.9\columnwidth]{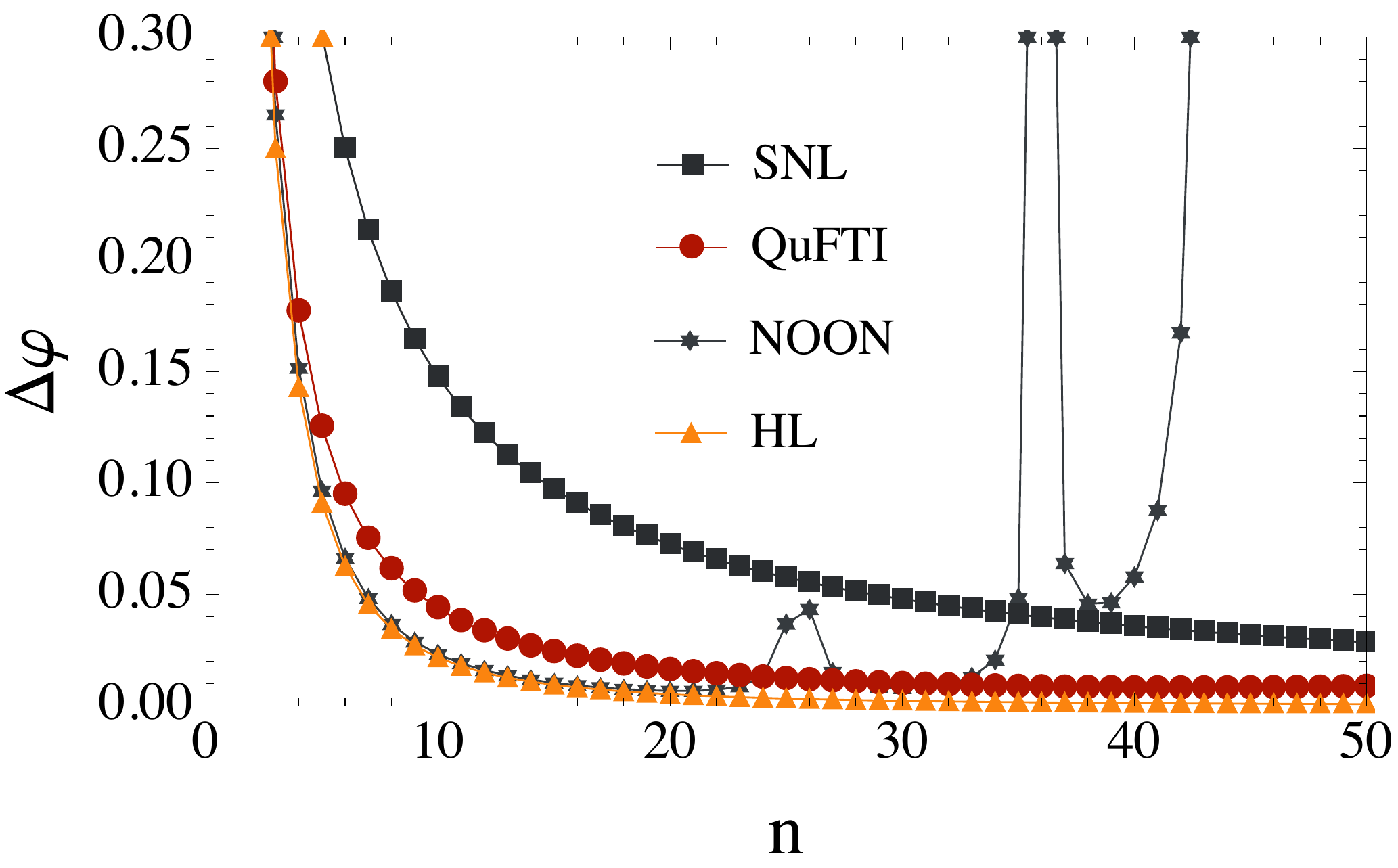}
\caption{The effect of dephasing on the NOON state and QuFTI where $\varphi=0.01, \chi=0.005$.  The NOON state is plotted with respect to $N$ for fair resource counting.} \label{fig:dephasingNOON}
\end{figure}

To model dephasing we investigate a random phase shift $\Delta\chi$ added to each mode separately. $\Delta\chi$ is a Gaussian random variable of zero mean but nonzero second order moment. The phase shift in the $j$th mode then becomes,
\begin{eqnarray}
e^{\pm i j \varphi}&\rightarrow& e^{\pm i j (\varphi + \Delta\chi)} \nonumber \\
&=& e^{\pm i j \varphi}e^{\pm i j \Delta\chi} \nonumber \\
&=& e^{\pm i j \varphi}\left(1 \pm i j\Delta\chi - \frac{1}{2}j\Delta\chi^2\pm\dots\right).
\end{eqnarray}
Using $\expec{\Delta\chi}=0$, $\expec{\Delta\chi^2}\neq0$, and that $\Delta\chi \ll \phi$ we simplify this to be, 
\begin{eqnarray}
e^{\pm i j \varphi}&\rightarrow& e^{\pm i j \varphi}\left(1- \frac{1}{2}j\Delta\chi^2 \pm\dots\right) \nonumber \\
&\approx& e^{\pm i j \varphi}e^{-\frac{1}{2} j^2 \Delta\chi^2}.
\end{eqnarray}

The signal $P$ in Eq. 10 from our work then changes in the presence of dephasing. The dependence that $P$ has on the unknown phase $\varphi$ does not depend on the mode number $j$. Then the term that depends on $\varphi$ becomes,
\begin{eqnarray}
\mathrm{cos}(n\phi) &=& \frac{1}{2}\left(e^{i n \varphi}+e^{-i n \varphi}\right) \nonumber \\
&\rightarrow& \frac{1}{2}\left(e^{i n \phi}+e^{- i n \phi}\right)e^{-\frac{�1}{2}n^2\Delta\chi^2} \nonumber \\
&=& \mathrm{cos}(n\phi)e^{-\frac{�1}{2}n^2\Delta\chi^2}
\end{eqnarray}
Using this substitution $P$ becomes,
\begin{eqnarray} \label{}
P &=& \Big|\mathrm{Per}(\hat{U}^{(n)})\Big|^2 \nonumber \\
&=& \frac{1}{n^{2n-2}}\prod_{j=1}^{n-1} \Big[a_n(j)\mathrm{cos}(n\phi)e^{-\frac{�1}{2}n^2\Delta\chi^2}+b_n(j) \Big].
\end{eqnarray} 
The factor $e^{-\frac{1}{2}n^2\Delta\chi^2}$ can be absorbed into $a_n(j)$ so that the derivation of $|\frac{\partial P}{\partial\phi}|$ in Eq. \ref{eq:dP} is identical.
Using this result we numerically plot the phase sensitivity with dephasing in Fig. \ref{fig:dephasing}.

In order to meaningfully analyze the dephased sensitivity, we would like to compare with other well known metrological schemes.  In Fig. \ref{fig:dephasingNOON}, we compare the QuFTI to the NOON state (with $N$ input photons for a fair resource comparison) and see that the QuFTI is far more robust against dephasing.

\bibliography{bibliography}

\begin{thebibliography}{44}
\expandafter\ifx\csname natexlab\endcsname\relax\def\natexlab#1{#1}\fi
\expandafter\ifx\csname bibnamefont\endcsname\relax
  \def\bibnamefont#1{#1}\fi
\expandafter\ifx\csname bibfnamefont\endcsname\relax
  \def\bibfnamefont#1{#1}\fi
\expandafter\ifx\csname citenamefont\endcsname\relax
  \def\citenamefont#1{#1}\fi
\expandafter\ifx\csname url\endcsname\relax
  \def\url#1{\texttt{#1}}\fi
\expandafter\ifx\csname urlprefix\endcsname\relax\def\urlprefix{URL }\fi
\providecommand{\bibinfo}[2]{#2}
\providecommand{\eprint}[2][]{\url{#2}}

\bibitem[{\citenamefont{Yurke}(1986)}]{bib:yurke1986input}
\bibinfo{author}{\bibfnamefont{B.}~\bibnamefont{Yurke}},
  \bibinfo{journal}{Phys. Rev. Lett} \textbf{\bibinfo{volume}{56}},
  \bibinfo{pages}{1515} (\bibinfo{year}{1986}).

\bibitem[{\citenamefont{Yuen}(1986)}]{bib:yuen1986generation}
\bibinfo{author}{\bibfnamefont{H.~P.} \bibnamefont{Yuen}},
  \bibinfo{journal}{Phys. Rev. Lett.} \textbf{\bibinfo{volume}{56}},
  \bibinfo{pages}{2176} (\bibinfo{year}{1986}).

\bibitem[{\citenamefont{Dowling}(1998)}]{bib:dowling1998correlated}
\bibinfo{author}{\bibfnamefont{J.~P.} \bibnamefont{Dowling}},
  \bibinfo{journal}{Phys. Rev. A} \textbf{\bibinfo{volume}{57}},
  \bibinfo{pages}{4736} (\bibinfo{year}{1998}).

\bibitem[{\citenamefont{Yurtsever et~al.}(2003)\citenamefont{Yurtsever,
  Strekalov, and Dowling}}]{bib:yurtsever2003interferometry}
\bibinfo{author}{\bibfnamefont{U.}~\bibnamefont{Yurtsever}},
  \bibinfo{author}{\bibfnamefont{D.}~\bibnamefont{Strekalov}},
  \bibnamefont{and} \bibinfo{author}{\bibfnamefont{J.}~\bibnamefont{Dowling}},
  \bibinfo{journal}{The Euro. Phys. J. D-Atomic, Molecular, Optical and Plasma
  Physics} \textbf{\bibinfo{volume}{22}}, \bibinfo{pages}{365}
  (\bibinfo{year}{2003}).

\bibitem[{\citenamefont{Nasr et~al.}(2003)\citenamefont{Nasr, Saleh, Sergienko,
  and Teich}}]{bib:nasr2003demonstration}
\bibinfo{author}{\bibfnamefont{M.~B.} \bibnamefont{Nasr}},
  \bibinfo{author}{\bibfnamefont{B.~E.} \bibnamefont{Saleh}},
  \bibinfo{author}{\bibfnamefont{A.~V.} \bibnamefont{Sergienko}},
  \bibnamefont{and} \bibinfo{author}{\bibfnamefont{M.~C.} \bibnamefont{Teich}},
  \bibinfo{journal}{Phys. Rev. Lett.} \textbf{\bibinfo{volume}{91}},
  \bibinfo{pages}{083601} (\bibinfo{year}{2003}).

\bibitem[{\citenamefont{Toussaint et~al.}(2004)\citenamefont{Toussaint, Jr.,
  Giuseppe, Bycenski, Sergienko, Saleh, and Teich}}]{bib:toussaint2004}
\bibinfo{author}{\bibfnamefont{K.}~\bibnamefont{Toussaint}},
  \bibinfo{author}{\bibnamefont{Jr.}}, \bibinfo{author}{\bibfnamefont{G.~D.}
  \bibnamefont{Giuseppe}}, \bibinfo{author}{\bibfnamefont{K.~J.}
  \bibnamefont{Bycenski}}, \bibinfo{author}{\bibfnamefont{A.~V.}
  \bibnamefont{Sergienko}}, \bibinfo{author}{\bibfnamefont{B.~E.~A.}
  \bibnamefont{Saleh}}, \bibnamefont{and} \bibinfo{author}{\bibfnamefont{M.~C.}
  \bibnamefont{Teich}}, \bibinfo{journal}{Physical Review A}
  \textbf{\bibinfo{volume}{70}}, \bibinfo{pages}{023801}
  (\bibinfo{year}{2004}).

\bibitem[{\citenamefont{Jones et~al.}(2009)\citenamefont{Jones, Karlen,
  Fitzsimons, Ardavan, Benjamin, Briggs, and Morton}}]{bib:jones2009magnetic}
\bibinfo{author}{\bibfnamefont{J.~A.} \bibnamefont{Jones}},
  \bibinfo{author}{\bibfnamefont{S.~D.} \bibnamefont{Karlen}},
  \bibinfo{author}{\bibfnamefont{J.}~\bibnamefont{Fitzsimons}},
  \bibinfo{author}{\bibfnamefont{A.}~\bibnamefont{Ardavan}},
  \bibinfo{author}{\bibfnamefont{S.~C.} \bibnamefont{Benjamin}},
  \bibinfo{author}{\bibfnamefont{G.~A.~D.} \bibnamefont{Briggs}},
  \bibnamefont{and} \bibinfo{author}{\bibfnamefont{J.~J.}
  \bibnamefont{Morton}}, \bibinfo{journal}{Science}
  \textbf{\bibinfo{volume}{324}}, \bibinfo{pages}{1166} (\bibinfo{year}{2009}).

\bibitem[{\citenamefont{Crespi et~al.}(2012)\citenamefont{Crespi, Lobino,
  Matthews, Politi, Neal, Ramponi, Osellame, and
  O'Brien}}]{bib:crespi2012measuring}
\bibinfo{author}{\bibfnamefont{A.}~\bibnamefont{Crespi}},
  \bibinfo{author}{\bibfnamefont{M.}~\bibnamefont{Lobino}},
  \bibinfo{author}{\bibfnamefont{J.~C.} \bibnamefont{Matthews}},
  \bibinfo{author}{\bibfnamefont{A.}~\bibnamefont{Politi}},
  \bibinfo{author}{\bibfnamefont{C.~R.} \bibnamefont{Neal}},
  \bibinfo{author}{\bibfnamefont{R.}~\bibnamefont{Ramponi}},
  \bibinfo{author}{\bibfnamefont{R.}~\bibnamefont{Osellame}}, \bibnamefont{and}
  \bibinfo{author}{\bibfnamefont{J.~L.} \bibnamefont{O'Brien}},
  \bibinfo{journal}{App. Phys. Lett.} \textbf{\bibinfo{volume}{100}},
  \bibinfo{pages}{233704} (\bibinfo{year}{2012}).

\bibitem[{\citenamefont{Rozema et~al.}(2014)\citenamefont{Rozema, Bateman,
  Mahler, Okamoto, Feizpour, Hayat, and Steinberg}}]{bib:rozema2014scalable}
\bibinfo{author}{\bibfnamefont{L.~A.} \bibnamefont{Rozema}},
  \bibinfo{author}{\bibfnamefont{J.~D.} \bibnamefont{Bateman}},
  \bibinfo{author}{\bibfnamefont{D.~H.} \bibnamefont{Mahler}},
  \bibinfo{author}{\bibfnamefont{R.}~\bibnamefont{Okamoto}},
  \bibinfo{author}{\bibfnamefont{A.}~\bibnamefont{Feizpour}},
  \bibinfo{author}{\bibfnamefont{A.}~\bibnamefont{Hayat}}, \bibnamefont{and}
  \bibinfo{author}{\bibfnamefont{A.~M.} \bibnamefont{Steinberg}},
  \bibinfo{journal}{Phys. Rev. Lett.} \textbf{\bibinfo{volume}{112}},
  \bibinfo{pages}{223602} (\bibinfo{year}{2014}).

\bibitem[{\citenamefont{Israel et~al.}(2014)\citenamefont{Israel, Rosen, and
  Silberberg}}]{bib:israel2014supersensitive}
\bibinfo{author}{\bibfnamefont{Y.}~\bibnamefont{Israel}},
  \bibinfo{author}{\bibfnamefont{S.}~\bibnamefont{Rosen}}, \bibnamefont{and}
  \bibinfo{author}{\bibfnamefont{Y.}~\bibnamefont{Silberberg}},
  \bibinfo{journal}{Phys. Rev. Lett.} \textbf{\bibinfo{volume}{112}},
  \bibinfo{pages}{103604} (\bibinfo{year}{2014}).

\bibitem[{\citenamefont{Holland and
  Burnett}(1993)}]{bib:holland1993interferometric}
\bibinfo{author}{\bibfnamefont{M.}~\bibnamefont{Holland}} \bibnamefont{and}
  \bibinfo{author}{\bibfnamefont{K.}~\bibnamefont{Burnett}},
  \bibinfo{journal}{Phys. Rev. Lett.} \textbf{\bibinfo{volume}{71}},
  \bibinfo{pages}{1355} (\bibinfo{year}{1993}).

\bibitem[{\citenamefont{Lee et~al.}(2002{\natexlab{a}})\citenamefont{Lee, Kok,
  and Dowling}}]{bib:lee2002quantum}
\bibinfo{author}{\bibfnamefont{H.}~\bibnamefont{Lee}},
  \bibinfo{author}{\bibfnamefont{P.}~\bibnamefont{Kok}}, \bibnamefont{and}
  \bibinfo{author}{\bibfnamefont{J.~P.} \bibnamefont{Dowling}},
  \bibinfo{journal}{J. Mod. Opt.} \textbf{\bibinfo{volume}{49}},
  \bibinfo{pages}{2325} (\bibinfo{year}{2002}{\natexlab{a}}).

\bibitem[{\citenamefont{Durkin and Dowling}(2007)}]{bib:durkin2007local}
\bibinfo{author}{\bibfnamefont{G.~A.} \bibnamefont{Durkin}} \bibnamefont{and}
  \bibinfo{author}{\bibfnamefont{J.~P.} \bibnamefont{Dowling}},
  \bibinfo{journal}{Phys. Rev. Lett.} \textbf{\bibinfo{volume}{99}},
  \bibinfo{pages}{070801} (\bibinfo{year}{2007}).

\bibitem[{\citenamefont{Dowling}(2008)}]{bib:dowling2008quantum}
\bibinfo{author}{\bibfnamefont{J.~P.} \bibnamefont{Dowling}},
  \bibinfo{journal}{Contemp. Phys.} \textbf{\bibinfo{volume}{49}},
  \bibinfo{pages}{125} (\bibinfo{year}{2008}).

\bibitem[{\citenamefont{Gerry and Campos}(2001)}]{bib:gerry2001generation}
\bibinfo{author}{\bibfnamefont{C.~C.} \bibnamefont{Gerry}} \bibnamefont{and}
  \bibinfo{author}{\bibfnamefont{R.}~\bibnamefont{Campos}},
  \bibinfo{journal}{Phys. Rev. A} \textbf{\bibinfo{volume}{64}},
  \bibinfo{pages}{063814} (\bibinfo{year}{2001}).

\bibitem[{\citenamefont{Kapale and
  Dowling}(2007)}]{bib:kapale2007bootstrapping}
\bibinfo{author}{\bibfnamefont{K.~T.} \bibnamefont{Kapale}} \bibnamefont{and}
  \bibinfo{author}{\bibfnamefont{J.~P.} \bibnamefont{Dowling}},
  \bibinfo{journal}{Phys. Rev. Lett.} \textbf{\bibinfo{volume}{99}},
  \bibinfo{pages}{053602} (\bibinfo{year}{2007}).

\bibitem[{\citenamefont{Lee et~al.}(2002{\natexlab{b}})\citenamefont{Lee, Kok,
  Cerf, and Dowling}}]{bib:lee2002linear}
\bibinfo{author}{\bibfnamefont{H.}~\bibnamefont{Lee}},
  \bibinfo{author}{\bibfnamefont{P.}~\bibnamefont{Kok}},
  \bibinfo{author}{\bibfnamefont{N.~J.} \bibnamefont{Cerf}}, \bibnamefont{and}
  \bibinfo{author}{\bibfnamefont{J.~P.} \bibnamefont{Dowling}},
  \bibinfo{journal}{Phys. Rev. A} \textbf{\bibinfo{volume}{65}},
  \bibinfo{pages}{030101} (\bibinfo{year}{2002}{\natexlab{b}}).

\bibitem[{\citenamefont{VanMeter et~al.}(2007)\citenamefont{VanMeter,
  Lougovski, Uskov, Kieling, Eisert, and Dowling}}]{bib:vanmeter2007general}
\bibinfo{author}{\bibfnamefont{N.}~\bibnamefont{VanMeter}},
  \bibinfo{author}{\bibfnamefont{P.}~\bibnamefont{Lougovski}},
  \bibinfo{author}{\bibfnamefont{D.}~\bibnamefont{Uskov}},
  \bibinfo{author}{\bibfnamefont{K.}~\bibnamefont{Kieling}},
  \bibinfo{author}{\bibfnamefont{J.}~\bibnamefont{Eisert}}, \bibnamefont{and}
  \bibinfo{author}{\bibfnamefont{J.~P.} \bibnamefont{Dowling}},
  \bibinfo{journal}{Phys. Rev. A} \textbf{\bibinfo{volume}{76}},
  \bibinfo{pages}{063808} (\bibinfo{year}{2007}).

\bibitem[{\citenamefont{Cable and Dowling}(2007)}]{bib:cable2007efficient}
\bibinfo{author}{\bibfnamefont{H.}~\bibnamefont{Cable}} \bibnamefont{and}
  \bibinfo{author}{\bibfnamefont{J.~P.} \bibnamefont{Dowling}},
  \bibinfo{journal}{Physical review letters} \textbf{\bibinfo{volume}{99}},
  \bibinfo{pages}{163604} (\bibinfo{year}{2007}).

\bibitem[{\citenamefont{Kok et~al.}(2007)\citenamefont{Kok, Munro, Nemoto,
  Ralph, Dowling, and Milburn}}]{bib:kok2007linear}
\bibinfo{author}{\bibfnamefont{P.}~\bibnamefont{Kok}},
  \bibinfo{author}{\bibfnamefont{W.~J.} \bibnamefont{Munro}},
  \bibinfo{author}{\bibfnamefont{K.}~\bibnamefont{Nemoto}},
  \bibinfo{author}{\bibfnamefont{T.~C.} \bibnamefont{Ralph}},
  \bibinfo{author}{\bibfnamefont{J.~P.} \bibnamefont{Dowling}},
  \bibnamefont{and} \bibinfo{author}{\bibfnamefont{G.}~\bibnamefont{Milburn}},
  \bibinfo{journal}{Rev. Mod. Phys.} \textbf{\bibinfo{volume}{79}},
  \bibinfo{pages}{135} (\bibinfo{year}{2007}).

\bibitem[{\citenamefont{Seshadreesan et~al.}(2013)\citenamefont{Seshadreesan,
  Kim, Dowling, and Lee}}]{bib:seshadreesan2013phase}
\bibinfo{author}{\bibfnamefont{K.~P.} \bibnamefont{Seshadreesan}},
  \bibinfo{author}{\bibfnamefont{S.}~\bibnamefont{Kim}},
  \bibinfo{author}{\bibfnamefont{J.~P.} \bibnamefont{Dowling}},
  \bibnamefont{and} \bibinfo{author}{\bibfnamefont{H.}~\bibnamefont{Lee}},
  \bibinfo{journal}{Phys. Rev. A} \textbf{\bibinfo{volume}{87}},
  \bibinfo{pages}{043833} (\bibinfo{year}{2013}).

\bibitem[{\citenamefont{Mayer et~al.}(2011)\citenamefont{Mayer, Tichy, Mintert,
  Konrad, and Buchleitner}}]{bib:mayer2011counting}
\bibinfo{author}{\bibfnamefont{K.}~\bibnamefont{Mayer}},
  \bibinfo{author}{\bibfnamefont{M.~C.} \bibnamefont{Tichy}},
  \bibinfo{author}{\bibfnamefont{F.}~\bibnamefont{Mintert}},
  \bibinfo{author}{\bibfnamefont{T.}~\bibnamefont{Konrad}}, \bibnamefont{and}
  \bibinfo{author}{\bibfnamefont{A.}~\bibnamefont{Buchleitner}},
  \bibinfo{journal}{Phys. Rev. A} \textbf{\bibinfo{volume}{83}},
  \bibinfo{pages}{062307} (\bibinfo{year}{2011}).

\bibitem[{\citenamefont{Gard et~al.}(2013)\citenamefont{Gard, Cross, Anisimov,
  Lee, and Dowling}}]{bib:gard2013quantum}
\bibinfo{author}{\bibfnamefont{B.~T.} \bibnamefont{Gard}},
  \bibinfo{author}{\bibfnamefont{R.~M.} \bibnamefont{Cross}},
  \bibinfo{author}{\bibfnamefont{P.~M.} \bibnamefont{Anisimov}},
  \bibinfo{author}{\bibfnamefont{H.}~\bibnamefont{Lee}}, \bibnamefont{and}
  \bibinfo{author}{\bibfnamefont{J.~P.} \bibnamefont{Dowling}},
  \bibinfo{journal}{JOSA B} \textbf{\bibinfo{volume}{30}},
  \bibinfo{pages}{1538} (\bibinfo{year}{2013}).

\bibitem[{\citenamefont{Gard et~al.}(2014{\natexlab{a}})\citenamefont{Gard,
  Olson, Cross, Kim, Lee, and Dowling}}]{bib:gard2014inefficiency}
\bibinfo{author}{\bibfnamefont{B.~T.} \bibnamefont{Gard}},
  \bibinfo{author}{\bibfnamefont{J.~P.} \bibnamefont{Olson}},
  \bibinfo{author}{\bibfnamefont{R.~M.} \bibnamefont{Cross}},
  \bibinfo{author}{\bibfnamefont{M.~B.} \bibnamefont{Kim}},
  \bibinfo{author}{\bibfnamefont{H.}~\bibnamefont{Lee}}, \bibnamefont{and}
  \bibinfo{author}{\bibfnamefont{J.~P.} \bibnamefont{Dowling}},
  \bibinfo{journal}{Phys. Rev. A} \textbf{\bibinfo{volume}{89}},
  \bibinfo{pages}{022328} (\bibinfo{year}{2014}{\natexlab{a}}).

\bibitem[{\citenamefont{Aaronson and Arkhipov}(2011)}]{bib:AaronsonArkhipov10}
\bibinfo{author}{\bibfnamefont{S.}~\bibnamefont{Aaronson}} \bibnamefont{and}
  \bibinfo{author}{\bibfnamefont{A.}~\bibnamefont{Arkhipov}},
  \bibinfo{journal}{Proc. ACM STOC (New York)} p. \bibinfo{pages}{333}
  (\bibinfo{year}{2011}).

\bibitem[{\citenamefont{Gard et~al.}(2014{\natexlab{b}})\citenamefont{Gard,
  Motes, Olson, Rohde, and Dowling}}]{bib:Chapter}
\bibinfo{author}{\bibfnamefont{B.~T.} \bibnamefont{Gard}},
  \bibinfo{author}{\bibfnamefont{K.~R.} \bibnamefont{Motes}},
  \bibinfo{author}{\bibfnamefont{J.~P.} \bibnamefont{Olson}},
  \bibinfo{author}{\bibfnamefont{P.~P.} \bibnamefont{Rohde}}, \bibnamefont{and}
  \bibinfo{author}{\bibfnamefont{J.~P.} \bibnamefont{Dowling}}
  (\bibinfo{year}{2014}{\natexlab{b}}), \eprint{arXiv:1406.6767}.

\bibitem[{See()}]{See}
See, \bibinfo{note}{for another practical application of the complexity of
  linear optics, J. Huh, G. G. Guerreschi, B. Peropadre, J. R. McClean, and A.
  Aspuru-Guzikl (2014), quant-ph/1412.8427.}

\bibitem[{\citenamefont{Lapaire et~al.}(2003)\citenamefont{Lapaire, Kok,
  Dowling, and Sipe}}]{bib:lapaire2003conditional}
\bibinfo{author}{\bibfnamefont{G.}~\bibnamefont{Lapaire}},
  \bibinfo{author}{\bibfnamefont{P.}~\bibnamefont{Kok}},
  \bibinfo{author}{\bibfnamefont{J.~P.} \bibnamefont{Dowling}},
  \bibnamefont{and} \bibinfo{author}{\bibfnamefont{J.}~\bibnamefont{Sipe}},
  \bibinfo{journal}{Physical Review A} \textbf{\bibinfo{volume}{68}},
  \bibinfo{pages}{042314} (\bibinfo{year}{2003}).

\bibitem[{\citenamefont{Knill et~al.}(2001)\citenamefont{Knill, Laflamme, and
  Milburn}}]{bib:LOQC}
\bibinfo{author}{\bibfnamefont{E.}~\bibnamefont{Knill}},
  \bibinfo{author}{\bibfnamefont{R.}~\bibnamefont{Laflamme}}, \bibnamefont{and}
  \bibinfo{author}{\bibfnamefont{G.~J.} \bibnamefont{Milburn}},
  \bibinfo{journal}{Nature} \textbf{\bibinfo{volume}{409}}, \bibinfo{pages}{46}
  (\bibinfo{year}{2001}).

\bibitem[{\citenamefont{Matthews et~al.}(2011)\citenamefont{Matthews, Politi,
  Bonneau, and O'Brien}}]{bib:matthews2011heralding}
\bibinfo{author}{\bibfnamefont{J.~C.} \bibnamefont{Matthews}},
  \bibinfo{author}{\bibfnamefont{A.}~\bibnamefont{Politi}},
  \bibinfo{author}{\bibfnamefont{D.}~\bibnamefont{Bonneau}}, \bibnamefont{and}
  \bibinfo{author}{\bibfnamefont{J.~L.} \bibnamefont{O'Brien}},
  \bibinfo{journal}{Phys. Rev. Lett.} \textbf{\bibinfo{volume}{107}},
  \bibinfo{pages}{163602} (\bibinfo{year}{2011}).

\bibitem[{\citenamefont{Spring et~al.}(2013)\citenamefont{Spring, Metcalf,
  Humphreys, Kolthammer, Jin, Barbieri, Datta, Thomas-Peter, Langford, Kundys
  et~al.}}]{bib:spring2013boson}
\bibinfo{author}{\bibfnamefont{J.~B.} \bibnamefont{Spring}},
  \bibinfo{author}{\bibfnamefont{B.~J.} \bibnamefont{Metcalf}},
  \bibinfo{author}{\bibfnamefont{P.~C.} \bibnamefont{Humphreys}},
  \bibinfo{author}{\bibfnamefont{W.~S.} \bibnamefont{Kolthammer}},
  \bibinfo{author}{\bibfnamefont{X.-M.} \bibnamefont{Jin}},
  \bibinfo{author}{\bibfnamefont{M.}~\bibnamefont{Barbieri}},
  \bibinfo{author}{\bibfnamefont{A.}~\bibnamefont{Datta}},
  \bibinfo{author}{\bibfnamefont{N.}~\bibnamefont{Thomas-Peter}},
  \bibinfo{author}{\bibfnamefont{N.~K.} \bibnamefont{Langford}},
  \bibinfo{author}{\bibfnamefont{D.}~\bibnamefont{Kundys}},
  \bibnamefont{et~al.}, \bibinfo{journal}{Science}
  \textbf{\bibinfo{volume}{339}}, \bibinfo{pages}{798} (\bibinfo{year}{2013}).

\bibitem[{\citenamefont{Broome et~al.}(2013)\citenamefont{Broome, Fedrizzi,
  Rahimi-Keshari, Dove, Aaronson, Ralph, and White}}]{bib:Broome2012}
\bibinfo{author}{\bibfnamefont{M.~A.} \bibnamefont{Broome}},
  \bibinfo{author}{\bibfnamefont{A.}~\bibnamefont{Fedrizzi}},
  \bibinfo{author}{\bibfnamefont{S.}~\bibnamefont{Rahimi-Keshari}},
  \bibinfo{author}{\bibfnamefont{J.}~\bibnamefont{Dove}},
  \bibinfo{author}{\bibfnamefont{S.}~\bibnamefont{Aaronson}},
  \bibinfo{author}{\bibfnamefont{T.~C.} \bibnamefont{Ralph}}, \bibnamefont{and}
  \bibinfo{author}{\bibfnamefont{A.~G.} \bibnamefont{White}},
  \bibinfo{journal}{Science} \textbf{\bibinfo{volume}{339}},
  \bibinfo{pages}{6121} (\bibinfo{year}{2013}).

\bibitem[{\citenamefont{Crespi et~al.}(2013)\citenamefont{Crespi, Osellame,
  Ramponi, Brod, Galv{\~a}o, Spagnolo, Vitelli, Maiorino, Mataloni, and
  Sciarrino}}]{bib:crespi2013integrated}
\bibinfo{author}{\bibfnamefont{A.}~\bibnamefont{Crespi}},
  \bibinfo{author}{\bibfnamefont{R.}~\bibnamefont{Osellame}},
  \bibinfo{author}{\bibfnamefont{R.}~\bibnamefont{Ramponi}},
  \bibinfo{author}{\bibfnamefont{D.~J.} \bibnamefont{Brod}},
  \bibinfo{author}{\bibfnamefont{E.~F.} \bibnamefont{Galv{\~a}o}},
  \bibinfo{author}{\bibfnamefont{N.}~\bibnamefont{Spagnolo}},
  \bibinfo{author}{\bibfnamefont{C.}~\bibnamefont{Vitelli}},
  \bibinfo{author}{\bibfnamefont{E.}~\bibnamefont{Maiorino}},
  \bibinfo{author}{\bibfnamefont{P.}~\bibnamefont{Mataloni}}, \bibnamefont{and}
  \bibinfo{author}{\bibfnamefont{F.}~\bibnamefont{Sciarrino}},
  \bibinfo{journal}{Nature Phot.} \textbf{\bibinfo{volume}{7}},
  \bibinfo{pages}{545} (\bibinfo{year}{2013}).

\bibitem[{\citenamefont{Ralph}(2013)}]{bib:ralph2013quantum}
\bibinfo{author}{\bibfnamefont{T.}~\bibnamefont{Ralph}},
  \bibinfo{journal}{Nature Phot.} \textbf{\bibinfo{volume}{7}},
  \bibinfo{pages}{514} (\bibinfo{year}{2013}).

\bibitem[{\citenamefont{Motes et~al.}(2013)\citenamefont{Motes, Dowling, and
  Rohde}}]{bib:motes2013spontaneous}
\bibinfo{author}{\bibfnamefont{K.~R.} \bibnamefont{Motes}},
  \bibinfo{author}{\bibfnamefont{J.~P.} \bibnamefont{Dowling}},
  \bibnamefont{and} \bibinfo{author}{\bibfnamefont{P.~P.} \bibnamefont{Rohde}},
  \bibinfo{journal}{Phys. Rev. A} \textbf{\bibinfo{volume}{88}},
  \bibinfo{pages}{063822} (\bibinfo{year}{2013}).

\bibitem[{\citenamefont{Spagnolo et~al.}(2014)\citenamefont{Spagnolo, Vitelli,
  Bentivegna, Brod, Crespi, Flamini, Giacomini, Milani, Ramponi, Mataloni
  et~al.}}]{bib:spagnolo2014experimental}
\bibinfo{author}{\bibfnamefont{N.}~\bibnamefont{Spagnolo}},
  \bibinfo{author}{\bibfnamefont{C.}~\bibnamefont{Vitelli}},
  \bibinfo{author}{\bibfnamefont{M.}~\bibnamefont{Bentivegna}},
  \bibinfo{author}{\bibfnamefont{D.~J.} \bibnamefont{Brod}},
  \bibinfo{author}{\bibfnamefont{A.}~\bibnamefont{Crespi}},
  \bibinfo{author}{\bibfnamefont{F.}~\bibnamefont{Flamini}},
  \bibinfo{author}{\bibfnamefont{S.}~\bibnamefont{Giacomini}},
  \bibinfo{author}{\bibfnamefont{G.}~\bibnamefont{Milani}},
  \bibinfo{author}{\bibfnamefont{R.}~\bibnamefont{Ramponi}},
  \bibinfo{author}{\bibfnamefont{P.}~\bibnamefont{Mataloni}},
  \bibnamefont{et~al.}, \bibinfo{journal}{Nature Phot.}
  \textbf{\bibinfo{volume}{8}}, \bibinfo{pages}{615} (\bibinfo{year}{2014}).

\bibitem[{\citenamefont{Scully and Fleischhauer}(1992)}]{bib:scully1992high}
\bibinfo{author}{\bibfnamefont{M.~O.} \bibnamefont{Scully}} \bibnamefont{and}
  \bibinfo{author}{\bibfnamefont{M.}~\bibnamefont{Fleischhauer}},
  \bibinfo{journal}{Physical review letters} \textbf{\bibinfo{volume}{69}},
  \bibinfo{pages}{1360} (\bibinfo{year}{1992}).

\bibitem[{\citenamefont{Reck et~al.}(1994)\citenamefont{Reck, Zeilinger,
  Bernstein, and Bertani}}]{bib:Reck94}
\bibinfo{author}{\bibfnamefont{M.}~\bibnamefont{Reck}},
  \bibinfo{author}{\bibfnamefont{A.}~\bibnamefont{Zeilinger}},
  \bibinfo{author}{\bibfnamefont{H.~J.} \bibnamefont{Bernstein}},
  \bibnamefont{and} \bibinfo{author}{\bibfnamefont{P.}~\bibnamefont{Bertani}},
  \bibinfo{journal}{Phys. Rev. Lett.} \textbf{\bibinfo{volume}{73}},
  \bibinfo{pages}{58} (\bibinfo{year}{1994}).

\bibitem[{\citenamefont{Scheel}(2004)}]{bib:ScheelPerm}
\bibinfo{author}{\bibfnamefont{S.}~\bibnamefont{Scheel}}
  (\bibinfo{year}{2004}), \eprint{quant-ph/0406127}.

\bibitem[{\citenamefont{Scully and Dowling}(1993)}]{bib:scully1993quantum}
\bibinfo{author}{\bibfnamefont{M.~O.} \bibnamefont{Scully}} \bibnamefont{and}
  \bibinfo{author}{\bibfnamefont{J.~P.} \bibnamefont{Dowling}},
  \bibinfo{journal}{Physical Review A} \textbf{\bibinfo{volume}{48}},
  \bibinfo{pages}{3186} (\bibinfo{year}{1993}).

\bibitem[{\citenamefont{Fukuda et~al.}(2011)\citenamefont{Fukuda, Fujii,
  Numata, Amemiya, Yoshizawa, Tsuchida, Fujino, Ishii, Itatani, Inoue
  et~al.}}]{bib:fukuda2011titanium}
\bibinfo{author}{\bibfnamefont{D.}~\bibnamefont{Fukuda}},
  \bibinfo{author}{\bibfnamefont{G.}~\bibnamefont{Fujii}},
  \bibinfo{author}{\bibfnamefont{T.}~\bibnamefont{Numata}},
  \bibinfo{author}{\bibfnamefont{K.}~\bibnamefont{Amemiya}},
  \bibinfo{author}{\bibfnamefont{A.}~\bibnamefont{Yoshizawa}},
  \bibinfo{author}{\bibfnamefont{H.}~\bibnamefont{Tsuchida}},
  \bibinfo{author}{\bibfnamefont{H.}~\bibnamefont{Fujino}},
  \bibinfo{author}{\bibfnamefont{H.}~\bibnamefont{Ishii}},
  \bibinfo{author}{\bibfnamefont{T.}~\bibnamefont{Itatani}},
  \bibinfo{author}{\bibfnamefont{S.}~\bibnamefont{Inoue}},
  \bibnamefont{et~al.}, \bibinfo{journal}{Opt. Express}
  \textbf{\bibinfo{volume}{19}}, \bibinfo{pages}{870} (\bibinfo{year}{2011}).

\bibitem[{\citenamefont{Ngah et~al.}(2015)\citenamefont{Ngah, Alibart,
  Labonté, D'Auria, and Tanzilli}}]{bib:LPOR201400404}
\bibinfo{author}{\bibfnamefont{L.~A.} \bibnamefont{Ngah}},
  \bibinfo{author}{\bibfnamefont{O.}~\bibnamefont{Alibart}},
  \bibinfo{author}{\bibfnamefont{L.}~\bibnamefont{Labonté}},
  \bibinfo{author}{\bibfnamefont{V.}~\bibnamefont{D'Auria}}, \bibnamefont{and}
  \bibinfo{author}{\bibfnamefont{S.}~\bibnamefont{Tanzilli}},
  \bibinfo{journal}{Laser \& Photonics Reviews} \textbf{\bibinfo{volume}{9}},
  \bibinfo{pages}{L1} (\bibinfo{year}{2015}), ISSN \bibinfo{issn}{1863-8899}.

\bibitem[{\citenamefont{Maier et~al.}(2014)\citenamefont{Maier, Gold, Forchel,
  Gregersen, M{\o}rk, H\"{o}fling, Schneider, and Kamp}}]{bib:Maier14}
\bibinfo{author}{\bibfnamefont{S.}~\bibnamefont{Maier}},
  \bibinfo{author}{\bibfnamefont{P.}~\bibnamefont{Gold}},
  \bibinfo{author}{\bibfnamefont{A.}~\bibnamefont{Forchel}},
  \bibinfo{author}{\bibfnamefont{N.}~\bibnamefont{Gregersen}},
  \bibinfo{author}{\bibfnamefont{J.}~\bibnamefont{M{\o}rk}},
  \bibinfo{author}{\bibfnamefont{S.}~\bibnamefont{H\"{o}fling}},
  \bibinfo{author}{\bibfnamefont{C.}~\bibnamefont{Schneider}},
  \bibnamefont{and} \bibinfo{author}{\bibfnamefont{M.}~\bibnamefont{Kamp}},
  \bibinfo{journal}{Opt. Express} \textbf{\bibinfo{volume}{22}},
  \bibinfo{pages}{8136} (\bibinfo{year}{2014}).

\bibitem[{\citenamefont{B.~R.~Bardan}(2013)}]{bib:Bardhan2013}
\bibinfo{author}{\bibfnamefont{J.~D.} \bibnamefont{B.~R.~Bardan},
  \bibfnamefont{K.~Jiang}}, \bibinfo{journal}{Phys. Rev. A}
  \textbf{\bibinfo{volume}{88}}, \bibinfo{pages}{023857}
  (\bibinfo{year}{2013}).

\end{thebibliography}

\end{document}